\documentclass[two column, fleqn,10pt]{wlscirep}
\title{Unconventional superconductivity in Y$_5$Rh$_6$Sn$_{18}$ probed by muon spin relaxation}

\author[1,2*]{Amitava Bhattacharyya}
\author[1,2,+]{Devashibhai Adroja}
\author[3]{Naoki Kase}
\author[2]{Adrian Hillier}
\author[3]{Jun Akimitsu}
\author[2,4]{Andre Strydom}
\affil[1]{ISIS Facility, Rutherford Appleton Laboratory, Chilton, Didcot Oxon, OX11 0QX, UK}
\affil[2]{Highly Correlated Matter Research Group, Physics Department, University of Johannesburg, PO Box 524, Auckland Park 2006, South Africa}
\affil[3]{Department of Physics and Mathematics, Aoyama-Gakuin University, Fuchinobe 5-10-1, Sagamihara, Kanagawa 252-5258, Japan}
\affil[4]{Max Planck Institute for Chemical Physics of Solids, N\"othnitzerstr. 40, D-01187 Dresden, Germany}

\affil[*]{amitava.bhattacharyya@stfc.ac.uk}

\affil[+]{devashibhai.adroja@stfc.ac.uk}


\newcommand{\yrhsn}{Y$_5$Rh$_6$Sn$_{18}$}

\begin{abstract}
Conventional superconductors are robust diamagnets that expels magnetic fields through the Meissner effect. It would therefore be unexpected if a superconducting ground state would support spontaneous magnetics fields. Such broken time-reversal symmetry states have been suggested for the high$-$temperature superconductors, but their identification remains experimentally controversial. We present magnetization, heat capacity, zero field and transverse field muon spin relaxation experiments on the recently discovered caged type  superconductor \yrhsn\ ($T_{\bf c}$ = 3.0 K).  The electronic heat capacity of \yrhsn\ shows a $T^3$ dependence below $T_{\bf c}$ indicating an anisotropic superconducting gap with a point node. This result is in sharp contrast to that observed in the isostructural Lu$_5$Rh$_6$Sn$_{18}$ which is a strong coupling $s-$wave superconductor. The temperature dependence of the deduced superfluid density \yrhsn\ is consistent with a BCS $s-$wave gap function, while the zero-field muon spin relaxation measurements strongly evidences unconventional superconductivity through a spontaneous appearance of an internal magnetic field below the superconducting transition temperature, signifying that the superconducting state is categorized by the broken time-reversal symmetry. 
\end{abstract}

\begin{document}

\flushbottom
\maketitle
%
%
\thispagestyle{empty}

\section*{Introduction}

Although an enormous amount of progress has been made to understand the unconventional behaviors of the superconductors beyond the conventional BCS theory, many obstacles obstacles remain to be overcome~\cite{Bardeen, Sigrist}. BCS superconductors expel magnetic field through the Meissner effect. This is very rare phenomenon if a superconducting (SC) ground state would support spontaneous internal fields. Similar kind of broken symmetry states hitherto proposed for the high-temperature superconductors, but their identification remains experimentally questionable. Broken symmetry can revise the physics of a system and is causal to novel behavior, such as unconventional or magnetic mediated superconductivity. Recently it has been shown that spin-orbit coupling with locally broken symmetry enables a giant spin polarization in a semiconductor~\cite{Schaibley}. Time$-$reversal symmetry (TRS) breaking is exceptional and has been detected directly in only a few unconventional superconductors, e.g., Sr$_2$RuO$_4$~\cite{gm,jx}, (U;Th)Be$_{13}$~\cite{rhh} and UPt$_3$~\cite{gml}, PrPt$_4$Ge$_{12}$~\cite{am}, (Pr;La)(Os;Ru)$_4$Sb$_{12}$~\cite{ya}, LaNiGa$_2$~\cite{ad2},  LaNiC$_2$~\cite{ad1}, and Re$_6$Zr~\cite{rps}. Zero field muon spin relaxation (ZF$-\mu$SR) is a powerful technique  to search for TRS breaking fields below the superconducting transition temperature ($T_{\bf c}$).

\begin{figure}[ht]
\centering
\includegraphics[trim= 0cm 0cm 0cm 0cm, clip=true, totalheight=0.18\textheight, angle=180]{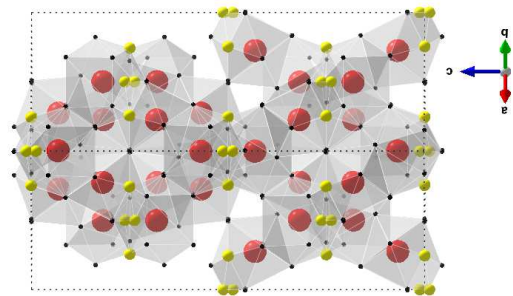}
\caption {(Color online) The tetragonal crystal structure of Y$_{5}$Rh$_{6}$Sn$_{18}$. The red spheres (largest) are Y, black spheres (smallest) are Sn, and the yellow spheres are Rh.}
\label{Fig:structure}
\end{figure}

Skutterudites (RT$_4$X$_{12}$), $\beta-$pyrochlore oxides (AOs$_2$O$_6$) and clathrates (Ge/Si-based) are materials classes with cage$-$like crystal structure have received considerable research interest in recent years and are the breeding ground of several unusual phenomena such as heavy fermion superconductivity, exciton$-$mediated superconducting state and Weyl fermions~\cite{zh}.  RT$_4$X$_{12}$ and RT$_{2}$X$_{20}$ exhibit a strong interplay between quadrupolar moment and superconductivity~\cite{kk,to}. The zero field (ZF)$-$muon spin relaxation ($\mu$SR) measurements in PrOs$_4$Sb$_{12}$ (first unconventional superconductor amongst Pr$-$ based metallic compounds) have revealed a significant upturn in the internal magnetic fields below the onset of superconductivity ($T_{\bf c}$ = 1.82 K)~\cite{Koga}. The low$-$lying crystal field excitations of Pr ions play a vital role in the superconductivity~\cite{Koga}. On the other hand, PrV$_2$Al$_{20}$ exhibits superconductivity below 50 mK in the antiferro$-$quadrupole ordered state, is an unusual specimen of a heavy-fermion superconductor based on strong hybridization among conduction electrons and nonmagnetic quadrupolar moments of the cubic $\Gamma_3$ ground doublet~\cite{Tsujimoto}. For PrV$_2$Al$_{20}$, the gapless mode associated with the quadrupole order is reflected in a cubic temperature dependence of electronic heat capacity. Non Fermi liquid behaviors are observed in PrIr$_2$Zn$_{20}$ and PrRh$_2$Zn$_{20}$ in the resistivity, specific heat and quadrupole ordering~\cite{Onimaru}. 

R$_5$Rh$_6$Sn$_{18}$ (R = Sc, Y, Lu) compounds also having a cage-like structure, crystallized in the tetragonal structure (see Fig.\ \ref{Fig:structure}) with the space group $I4_1/acd$ and Z = 8, where R occupies two sites of different symmetry ~\cite{sm},  exhibit superconductivity~\cite{jpr} below 5 K (Sc), 3 K (Y), and 4 K (Lu). The crystal structure is analogous to the skutterudite structure with a Pr$-$based heavy fermion superconductor~\cite{Bauer}. Lu$_5$Rh$_6$Sn$_{18}$ is a conventional BCS type superconductor, the gap structure of \yrhsn\ is found to be strongly anisotropic as exposed from the specific heat ($C_P$) measurements; the heat capacity coefficient ($\gamma$) of \yrhsn\ in the mixed state is found to follow a square root field dependence. This is a sign of anisotropic superconducting gap. The superconducting properties of \yrhsn\ thus have a correspondence with those of the anisotropic $s-$wave superconductor YNi$_2$B$_2$C except for the difference in superconducting transition temperature~\cite{kase2}.  On the other hand, the $\gamma$ of Lu$_5$Rh$_6$Sn$_{18}$ shows linear field dependence, which points to an isotropic superconducting gap. The unconventional superconductivity of nonmagnetic YNi$_2$B$_2$C draws considerable attention from several standpoints, such as high$-T_{\bf c}$ superconductivity among intermetallics~\cite{Mazumdar, Cava, Cava1} and anisotropic superconducting gap~\cite{Nohara, Nohara1}. Numerous measurements indicate that YNi$_2$B$_2$C has anisotropic superconducting gap (point-node type). K. Izawa $et. al.$ ~\cite{Izawa} suggest that the superconducting gap structure of YNi$_2$B$_2$C has point nodes located along the $a$ and $b$ axes by thermal conductivity measurements. For \yrhsn\ the field-angle-resolved specific heat $C_P(\phi)$ measurements in a rotating magnetic field $H$ exposes a clear fourfold angular oscillation of $C_P (\phi)$, signifying that the field$-$induced quasiparticle density of state is strongly anisotropic~\cite{nk2}. Additionally no considerable angular oscillation was observed in $C_P(\phi)$ of Lu$_5$Rh$_6$Sn$_{18}$, a reference compound of an isotropic superconducting gap.   

\begin{figure}[ht]
\centering
\includegraphics[width= \linewidth]{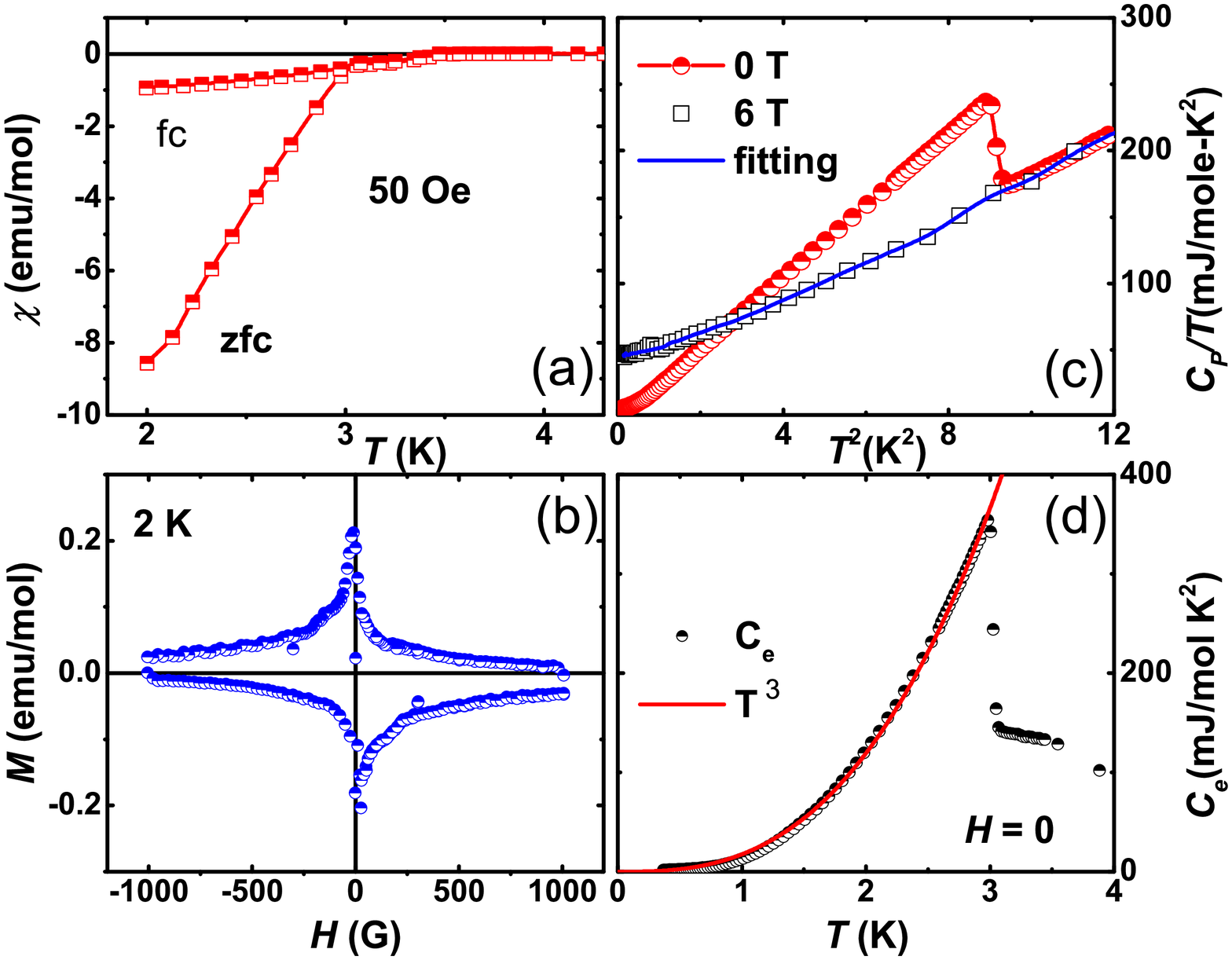}
\caption {(Color online) (a) Temperature dependence of the dc susceptibility $\chi(T)$ of Y$_{5}$Rh$_{6}$Sn$_{18}$ under the field of 50 Oe for zero field cooled (ZF) and field cooled (FC) sequences. (b) Low field part of the magnetization loop at 2.0 K. (c) shows the  $T^2$ dependence of the specific heat divided by $T$ .i.e. $C_P/T$ vs. $T^2$. The solid line shows the fit (using $C_P/T = \gamma +\beta T^2 + \delta T^4$). (d) temperature dependent of electronic specific heat $C_e$ under zero field after subtracting the lattice contribution for \yrhsn.}
\label{Fig:MTCP}
\end{figure}

We have recently reported superconducting properties of the caged type compound Lu$_5$Rh$_6$Sn$_{18}$ using magnetization, heat capacity, and muon spin relaxation measurements~\cite{Bhattacharyya1}. ZF$-\mu$SR measurements reveal the spontaneous appearance of an internal magnetic field below the superconducting transition temperature, which indicates that the superconducting state in this material is characterized by the broken time-reversal symmetry~\cite{Bhattacharyya1}.  From a series of experiments~\cite{kase2} on R$_5$Rh$_6$Sn$_{18}$ (R = Lu, Sc, Y and Tm), it was concluded that the superconducting gap structure is strongly dependent on the rare earth atom, but whose origin remains~\cite{nk2}. In this work, we address these matters by ZF$-\mu$SR measurements for the \yrhsn\ system. The results unambiguously reveal the spontaneous appearance of an internal magnetic field in the SC state, providing clear evidence for broken time reversal symmetry.

\section*{Results and Discussion}

The bulk nature of superconductivity in \yrhsn\ was established by the magnetic susceptibility $\chi(T)$, as shown in Fig.\ \ref{Fig:MTCP} (a). The low-field susceptibility displays a robust diamagnetic signal due to a superconducting transition at  $T_{\bf c}$ = 3.0 K. Fig.\ \ref{Fig:MTCP} (c) shows the magnetization $M(H)$ curve at 2 K, which is typical for type$-$II superconductivity. Remarkably, the electrical resistivity $\rho(T)$ of \yrhsn\ shows uncommon (not shown here) non-metallic behavior~\cite{nk2} on cooling down to just above $T_{\bf c}$ with a high residual resistivity $\rho_0$ of 350 $\mu\Omega$ cm. A moderately rare state in which the anisotropic superconductivity persists in an incoherent metallic state is suggested to occur in \yrhsn.

Fig.\ \ref{Fig:MTCP} (b) displays the $C_P(T)$  at $H$ = 0 and 6.0 T. Below 3.0 K in zero field a sharp anomaly is detected demonstrating the superconducting transition which matches well with $\chi(T)$ data. Subsequently the normal$-$state specific heat was found to be invariable under external magnetic fields. The normal-state electronic specific heat coefficient $\gamma$ and the lattice specific heat coefficient $\beta$ were deduced from the data in a field of 6.0 T by a least-square fitting of the $C_P/T$ data to $C_P/T = \gamma +\beta T^2 + \delta T^4$. This analysis provides a Sommerfeld coefficient $\gamma$ = 38.13(3) mJ/(mol-K$^2$) and the Debye temperature $\Theta_D$ = 183(2) K. We found the specific heat jump $\Delta C_P(T_{\bf c})$ = 222 mJ/(mol K) and $T_{\bf c}$ = 3.0 K, which yields  $\Delta C$/$\gamma T_{\bf c}$ = 1.96. 

\begin{figure}[ht]
\centering
\includegraphics[width = \linewidth]{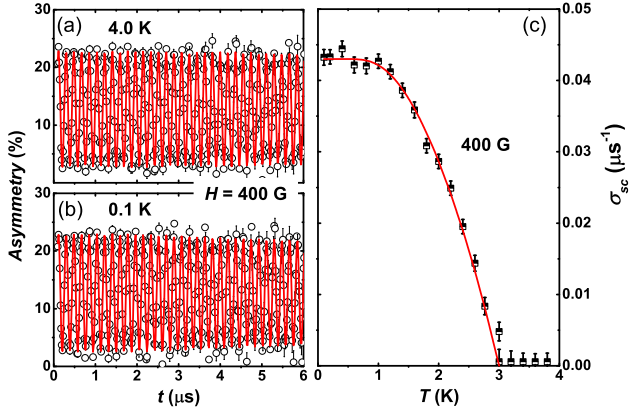}
\caption {(Color online) TF$-$field $\mu$SR spin precession signals of \yrhsn\ taken at applied magnetic field of $H$ = 400 G. Here we have shown only the real component: (a) in the normal state at 4.0 K and (b) in the superconducting state at 0.1 K. Solid lines represent fits to the data using Eq.\ \ref{eq:eq1}. (c) Temperature dependence of the muon Gaussian relaxation rate  $\sigma_{sc}(T)$. The line is a fit to the data using an isotropic model (Eq.\ \ref{eq:eq2}).}
\label{Fig:TFM}
\end{figure}

Fig.\ \ref{Fig:MTCP} (d) shows the electronic specific heat $C_e$ which was obtained after subtraction of the phonon part, to illuminate the superconducting gap symmetry. In case of line-nodes in the superconducting gap structure, $T^2$ dependence of the $C_e$ is anticipated below $T_{\bf c}$. We find a power law $C_e$ = c$T^\alpha$ with an exponent close to 3 ($\alpha$ = 2.93) over a comprehensive temperature region. This cubic temperature dependence suggest that \yrhsn\ has an anisotropic superconducting gap with a point node, such as has been found for the borocarbides YNi$_2$B$_2$C and LuNi$_2$B$_2$C. Analogous behavior is also detected in the heavy fermion superconductor UBe$_{13}$, which unveils a $T^{2.9}$ dependences of the specific heat below $T_{\bf c}$ together with a higher fraction~\cite{ott} of the ratio $\Delta C$/$\gamma T_{\bf c}$ = 2.5. On the other hand, $C_e$ of Lu$_5$Rh$_6$Sn$_{18}$ behaves as an exponential dependence. We obtained the specific heat jump $\Delta C_P(T_{\bf c})$ = 397$\pm$(3) mJ/(mol K) and $T_{\bf c}$ = 4.0$\pm$(0.2) K, which yields  $\Delta C$/$\gamma T_{\bf c}$ = 2.06$\pm$(0.03) for Lu$_5$Rh$_6$Sn$_{18}$. From the exponential dependence of $C_e$, we obtained 2$\Delta$(0)/$k_B$$T_{\bf c}$ to be 4.26$\pm$(0.04) for Lu$_5$Rh$_6$Sn$_{18}$. Because this value is relatively larger than that of the theoretical BCS limit of weak-coupling superconductor (3.54), Lu$_5$Rh$_6$Sn$_{18}$ compound is characterized as a strong-coupling superconductor~\cite{Bhattacharyya1, nk2}.

In order to determine the superfluid density or superconducting gap structure of \yrhsn\ we have carried out TF$-\mu$SR measurements at 400 G (well above lower critical field) applied magnetic field at 0.1 and 4.0 K as shown in Fig.\ \ref{Fig:TFM} (a) and (b). A clear difference is evident in the relaxation rate  below and above $T_{\bf c}$. Below $T_{\bf c}$, the TF$-\mu$SR precession signal decays with time due to inhomogeneous field distribution of the flux-line lattice. The analysis of our TF$-\mu$SR asymmetry spectra was carried out in the time domain using the following functional form,

\begin{align}
\begin{split}
G_{osc}(t) = A_0\cos(2\pi \nu_s t+\phi_1)\exp\left({-\frac{\sigma^2t^2}{2}}\right)\\ + A_{bg}\cos(2\pi \nu_{bg} t+\phi_2)
\end{split}
\label{eq:eq1}
\end{align}

where $\nu_s$ and $\nu_{bg}$ are the frequencies of the muon precession signal and background signal, respectively with phase angle $\phi_i$ ($i$ = 1, 2).  The first term gives the overall sample relaxation rate $\sigma$; there are contributions from both the vortex lattice ($\sigma_{sc}$) and nuclear dipole moments ($\sigma_{nm}$, which is expected to be constant over the whole temperature region) below $T_{\bf c}$ [where $\sigma$ = $\sqrt{(\sigma_{sc}^2+\sigma_{nm}^2)}$]. The contribution from the vortex lattice was determined by quadratically subtracting the background nuclear dipolar relaxation rate obtained from spectra measured above $T_{\bf c}$. The relaxation rate from the vortex lattice is directly associated to the magnetic penetration depth, the superconducting gap function can be demonstrated by,

\begin{align}
\frac{\sigma_{sc}(T)}{\sigma_{sc}(0)}=\frac{\lambda^{-2}(T)}{\lambda^{-2}(0)}=1+2\int_{\Delta(T)}^{\infty}\left (\frac{\delta f}{\delta E}\right) \frac{EdE}{\sqrt{E^2-\Delta(T)^2}}
\label{eq:eq2}
\end{align}

where 
\begin{align}
f = [1+exp(-E/k_B T)]^{-1}
\label{eq:eq4}
\end{align}

is the Fermi function~\cite{mt}. The $T$ dependence  of the superconducting  gap function is estimated by the expression~\cite{ac} 

\begin{align}
\delta(T/T_{\bf c})=\tanh\{1.82[1.018(T_{\bf c}/T-1)]^{0.51}\}
\label{eq:eq5}
\end{align}

Fig.\ \ref{Fig:TFM} (b) shows the $T$ dependence of the term $\sigma_{sc}$ which can be directly connected to the superfluid density. From this, the nature of the superconducting gap can be determined. The data can be well modeled using a single isotropic $s-$wave gap of 0.5$\pm$0.05 meV. This gives a gap of 2$\Delta$/$k_B$$T_{\bf c}$ = 3.91$\pm$0.03. For Lu$_5$Rh$_6$Sn$_{18}$, the analysis of  temperature dependence of the magnetic penetration depth suggest an isotropic $s-$wave character for the superconducting gap with a gap value 2$\Delta$/$k_B$$T_{\bf c}$ = 4.4$\pm$0.02, which is higher than the 3.53 expected for BCS superconductors. This is an indication of the strong electron$-$phonon coupling in the superconducting state. \yrhsn\ is a type II superconductor, supposing that approximately all the normal state carriers ($n_e$) contribute to the superconductivity (i.e. $n_s\sim n_e$), we have estimated the values of effective mass of the quasiparticles $m^*\approx 1.21 m_e$ and  superconducting electron density $\approx$ 2.3 $\times$10$^{28}$ m$^{-3}$ respectively. Additional details on these calculations can be found in Ref. \cite{adsd,vkasd,dtasd}.

\begin{figure}[ht]
\centering
\includegraphics[width = \linewidth]{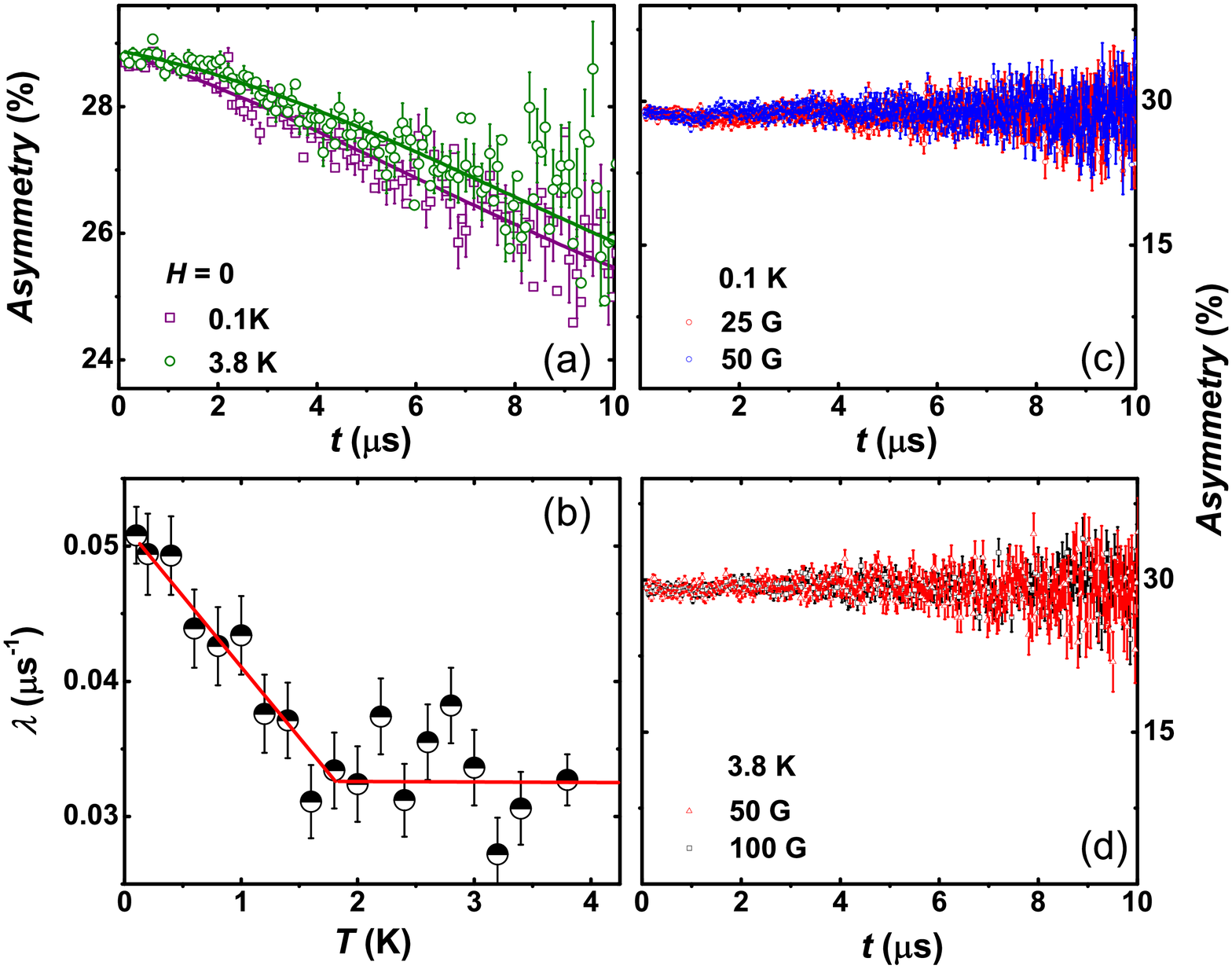}
\caption {(Color online) (a) Zero-field $\mu$SR time spectra for \yrhsn\ measured across the superconducting transition temperature 3 K, are shown together with lines that are least squares fits to the data using Eq.\ \ref{eq:eq3}. (b) shows the temperature dependence of the exponential relaxation rate measured in zero magnetic field of \yrhsn. The lines are guides to the eye. The extra relaxation below $T_{\bf c}$ indicates additional internal magnetic fields and, consequently, suggests the superconducting state has broken time reversal symmetry. Longitudinal field $\mu$SR time spectrum taken at  (c) 0.1 K and  (d) 3.8 K in presence of various applied magnetic fields.}
\label{Fig:ZFM}
\end{figure}

Fig.\ \ref{Fig:ZFM} (a) shows the time evolution of the zero field muon spin relaxation asymmetry spectra in \yrhsn\ at temperatures above and below $T_{\bf c}$. A constant background signal arising from muons stopping in the silver sample holder has been deducted from the data. Below $T_{\bf c}$, we observed that muon spin relaxation became faster with decreasing temperature down to lowest temperature, which is indicating the appearance of the spontaneous local field in the superconducting phase. No signature of muon spin precession is visible, which excludes internal fields produced by magnetic ordering as the origin of the spontaneous field. The only possibility is that the muon$-$spin relaxation is due to static, randomly oriented local fields associated with the nuclear moments at the muon site. The ZF$-\mu$SR spectra for \yrhsn\ can be well described by the damped Gaussian Kubo$-$Toyabe (K$-$T) function,

\begin{align}
G(t) =A_1 G_{KT}(t)\exp({-\lambda t})+A_{bg}
\label{eq:eq3}
\end{align}

where
\begin{align}
G_{KT}(t) =\left[\frac{1}{3}+\frac{2}{3}(1-\sigma_{KT}^2t^2)\exp\left({-{\frac{\sigma_{KT}^2t^2}{2}}}\right)\right]
\end{align}

is the K$-$T functional form expected from an isotropic Gaussian distribution of randomly oriented static (or quasistatic) local
fields at muon sites. $\lambda$ is the electronic relaxation rate, $A_0$ is the initial asymmetry, $A_{bg}$ is the background.  The parameters $\sigma_{KT}$, $A_0$, and $A_{bg}$ [not shown here] are found to be temperature independent all across the phase transition.

Fig.\ \ref{Fig:ZFM} (c) shows the temperature dependence of $\lambda$. In the normal state above $T_{\bf c}$,  $\lambda$ is due to dipolar fields from neighboring nuclear magnetic moments. It is extraordinary that $\lambda$ shows a substantial rise below $T_{\bf c}$, signifying the presence of a spontaneous internal field associated with the superconductivity. Further the temperature dependent $\lambda$ exhibits nearly linear increase with decreasing temperature below $T_{\bf c}$.  This observation provides unequivocal sign that TRS is broken in the SC state, below $T_{\bf c}$ that may suggest a possibility of an unusual superconducting mechanism below 2 K,  of \yrhsn. Such a change in $\lambda$ has only been detected in superconducting Sr$_2$RuO$_4$, LaNiC$_2$, and Lu$_5$Rh$_6$Sn$_{18}$~\cite{gm,ad1,Bhattacharyya1}.  This rise in $\lambda$ can be described in terms of a considerable internal field with a very small frequency as conferred by Luke {\it et. al.}~\cite{gm} for Sr$_2$RuO$_4$. This propose that the field distribution is Lorentzian in nature analogous to Sr$_2$RuO$_4$. The temperature dependence of $\lambda$ compares quantitatively with that in  in Sr$_2$RuO$_4$, Lu$_5$Rh$_6$Sn$_{18}$, LaNiC$_2$ and  \yrhsn\ and thus we attribute this behavior of $\lambda$ to the TRS breaking below $T_{\bf c}$ in \yrhsn.  A longitudinal magnetic field of just 25 G [Figs.\ \ref{Fig:ZFM} (b) and (d)] removes any relaxation due to the spontaneous fields and is sufficient to fully decouple the muons from this relaxation channel. This in turn shows that the associated magnetic fields are in fact static or quasistatic on the time scale of the muon precession. These observations further support the broken TRS in the superconducting state of \yrhsn. The increase in the exponential relaxation for Lu$_5$Rh$_6$Sn$_{18}$ below $T_{\bf c}$ is, 0.045 $\mu$s$^{-1}$, which corresponds to a characteristic field strength $\lambda/\gamma_\mu$=  0.5 G. For \yrhsn,  the rise in the exponential relaxation below $T_{\bf c}$ is, 0.0184 $\mu$s$^{-1}$, which resembles to a characteristic field strength $\lambda/\gamma_\mu$=  0.21 G, where $\gamma_\mu$ is the muon gyromagnetic ratio = 13.55 MHz/T. This is about half the value as observed in Lu$_5$Rh$_6$Sn$_{18}$, the B phase of UPt$_3$ and Sr$_2$RuO$_4$~\cite{gml}. No theoretical estimates of the characteristic fields strength in \yrhsn\ are yet presented; however, we presume them to be analogous to those in Sr$_2$RuO$_4$ and UPt$_3$ as the fields should arise from a analogous mechanism.

Time reversal symmetry breaking in the superconducting state has significant consequences for the symmetry of pairing and for the quasi-particle spectrum. A typical symmetry analysis~\cite{Sigrist1991} carried out under the theory of strong spin orbit coupling, yields two possibilities, one with singlet pairing ($d+id$ character) and additional one triplet pairing (non-unitary). The singlet pairing state has a line node and two point nodes, and the non-unitary triplet state has two point nodes. Below the superconducting transition temperature the thermodynamics of the singlet state would be dominated by the line node, yielding quadratic temperature dependence of the electronic specific heat. Likewise, the non-unitary triplet pairing state would be dominated by the point nodes, which happen to be shallow (a result protected by symmetry) and consequently also lead to $C_e \sim T^2$~\cite{Mazidian2013}.  Though, because of the location of the nodes in the triplet case, fully-gapped behavior may be recovered depending on the topology of the Fermi surface. Furthermore certain limiting cases of the triplet state correspond to regular, i.e. linear point nodes (cubic temperature dependence of the electronic heat capacity ) as well as to a more unusual state with a nodal surface (gapless superconductivity, $C_e \sim T$).  The allowed pairing states and their quasiparticle spectra are discussed in detail in the Supplementary Online Material~\cite{jorge}.

\section*{Conclusions}
In summary, we have investigated the vortex state in \yrhsn\ by using zero field and transverse field muon spin relaxation measurements. Below the superconducting transition temperature, the ZF-$\mu$SR measurements revealed an appreciable increase in the internal magnetic fields which does not coincide with the superconducting phase transition at $T_{\bf c}$ = 3.0 K, and which may indicates different nature of SC below and above 2 K. The appearance of spontaneous magnetic fields in our ZF-$\mu$SR measurements, deliver undoubted evidence  for TRS breaking in this material and an unconventional pairing mechanism. TF$-\mu$SR measurements yield a magnetic penetration depth that is exponentially flat at low temperatures, and so our data can be fit to a single-gap BCS model. Symmetry analysis suggests either a singlet $d+id$ state with a line node or, alternatively, nonunitary triplet pairing with point nodes, which may be linear or shallow and can become fully gapped depending on the Fermi surface topology~\cite{jorge}. The heat capacity below $T_{\bf c}$ exhibits $\sim T^3$ supporting point nodes scenario.

\section*{Methods}

An important step to study the intrinsic, particularly anisotropic properties is to grow sizable single crystals. The single crystals of \yrhsn\ were grown by liquefying the constituent elements in an excess of Sn$-$flux in the proportion of Y:Rh:Sn = 1:2:20. The quartz tube was heated up to 1323 K, kept at this temperature for about 3 h, and cooled down to 473 K at a rate of 5 $^\circ$C/h, taking 168 hr in total. The excess flux was detached from the crystals by spinning the ampoule in a centrifuge~\cite{jpr}. Several single crystals of \yrhsn\ were obtained having typical dimensions of 3$\times$3$\times$2 mm$^3$. Laue patterns were recorded with a Huber Laue diffractometer. Well defined Laue diffraction spots indicates the high quality of the single crystals. The phase purity was inferred from the powder x$-$ray  patterns, which were indexed as the \yrhsn\ phase with the space group~\cite{jpr} $I4_1/acd$ [lattice parameters: $a$ = 1.375(3) nm, $c$ = 2.745(1)]. The magnetization data were measured using a Quantum Design Superconducting Quantum Interference Device. The heat capacity were measured down to 500 mK using Quantum Design Physical Properties Measurement System equipped with a $^3$He refrigerator.

Muon spin relaxation is a dynamic method to resolve the type of the pairing symmetry in superconductors~\cite{js}. The mixed or vortex state in case of type-II superconductors gives rise to a spatial distribution of local magnetic fields; which demonstrate itself in the $\mu$SR signal through a relaxation of the muon polarization. We further employed the muon spin relaxation technique to examine the superconducting ground state. The zero field, transverse field (TF) and longitudinal field (LF) measurements were performed at the MUSR spectrometer at the ISIS Pulsed Neutron and Muon Facility located at the Rutherford Appleton Laboratory, United Kingdom. The ZF$-\mu$SR experiments were conducted in the longitudinal geometry. The unaligned single crystals were mounted on a high purity silver plate (99.995\%) using diluted GE Varnish for cryogenic heatsinking, which was placed in a dilution refrigerator with a temperature range of 100 mK to 4.4 K. TF$-\mu$SR experiments were performed in the superconducting mixed state in applied field of 400 G, well above the lower critical field of 18 G of \yrhsn. Data were collected in the field$-$cooled mode where the magnetic field was applied above the superconducting transition and the sample was then cooled down to base temperature. Using an active compensation system the stray magnetic fields at the sample position were canceled to a level of 0.01 G. Spin-polarized muon pulses were implanted into the sample and the positrons from the subsequent decay were collected in positions either forward or backwards of the initial muon spin direction. The asymmetry of the muon decay is calculated by, $G_z(t) =[ {N_F(t) -\alpha N_B(t)}]/[{N_F(t)+\alpha N_B(t)}]$, where $N_B(t)$ and $N_F(t)$ are the number of counts at the detectors in the forward and backward positions and $\alpha$ is a constant determined from calibration measurements made in the paramagnetic state with a small (20 G) applied transverse magnetic field. The data were analyzed using the free software package WiMBDA~\cite{FPW}.

\noindent 

\section*{Acknowledgements}

A.B. would like to acknowledge FRC of UJ, SA-NRF and ISIS-STFC for funding support. D.T.A. and A.D.H. would like to thank CMPC-STFC, grant number CMPC-09108, for financial support. A.M.S. thanks the SA-NRF (Grant 93549) and UJ Research Committee and the Faculty of Science for financial
support.

\section*{Author contributions statement}

Research problem was formulated by A.B., D.T.A. and A.M.S. The samples were prepared by N.K. and J.A. Experiments were carried out by A.B., A.D.H. A.M.S. and D.T.A. The data analysis and manuscript was prepared by A.B. All authors reviewed the manuscript.

\section*{Additional information}

Competing financial interests: The authors declare no competing financial interests.

{\bf How to cite this article}: Bhattacharyya, A. et al. Unconventional superconductivity in Y$_5$Rh$_6$Sn$_{18}$ probed by muon spin relaxation. Sci. Rep. 5, 12926; doi: 10.1038/srep12926 (2015).

\end{document}